\documentclass{PoS}

\title{
Dark matter and LHC: Complementarities\\ and limitations
}

\ShortTitle{
Dark matter and LHC: Complementarities and limitations
}

\author{\speaker{G. Robbins}$^{,1,2}$, F. Mahmoudi$^{1,2,3}$, A. Arbey$^{1,2,3}$, M. Boudaud$^{4}$\\
      ~\\$^1$  Univ Lyon, Univ Lyon 1, ENS de Lyon, CNRS, Centre de Recherche Astrophysique de Lyon UMR5574, F-69230 Saint-Genis-Laval, France
      \vspace*{0.2cm}\\
        $^2$ Univ Lyon, Univ Lyon 1, CNRS/IN2P3, Institut de Physique Nucl\'eaire de Lyon UMR5822, F-69622 Villeurbanne, France\vspace*{0.2cm}\\
       $^3$ CERN, Theoretical Physics Department, CH-1211 Geneva 23, Switzerland\vspace*{0.2cm}\\
        $^4$ Laboratoire de Physique Th\'eorique et Hautes \'Energies (LPTHE), UMR 7589 CNRS \& UPMC, 4 Place Jussieu, F-75252 Paris, France\vspace*{0.2cm}\\
         E-mail: \email{glenn.robbins@univ-lyon1.fr}
        }

\abstract{It is well known that dark matter density measurements, indirect and direct detection experiments, importantly complement the LHC in setting strong constraints on new physics scenarios. Yet, dark matter searches are subject to limitations which need to be considered for realistic analyses. For illustration, we explore the parameter space of the phenomenological MSSM and discuss the interplay of the constraints from dark matter searches and the LHC, and analyse the impact of the astrophysical uncertainties in some detail.}

\FullConference{EPS-HEP 2017, European Physical Society conference on High Energy Physics\\
		5-12 July 2017\\
		Venice, Italy}

\def\MED{{\sc Med }}
\def\MAX{{\sc Max }}

\begin{document}

\section{Introduction}
The Large Hadron Collider (LHC) is currently operating at 13 TeV with the hope of discovering hints of the existence of New Physics (NP). Supersymmetry (SUSY) is one of the most studied scenarios and the main focus of NP searches at the LHC. No new particle has been discovered so far, yet strong constraints on superparticle masses and interactions are obtained. One of the strongest motivations for SUSY is that it offers a good dark matter (DM) particle candidate, and it is therefore useful to study the complementarities between LHC and DM detection constraints. However, DM detection searches suffer from large astrophysical uncertainties, namely on the DM distribution in the Galaxy and on cosmic ray propagation through the galactic medium. In this study, we evaluated the effects of those uncertainties on direct and indirect DM detection experimental limits, and examined their impact on the constraints on the minimal supersymmetric extension of the Standard Model (MSSM) and the interplay of those constraints with collider limits. More precisely, we considered the phenomenological MSSM (pMSSM) with its 19 parameters, which is the most general R-parity and CP-conserving MSSM scenario respecting minimal flavour violation, and we assumed that the lightest neutralino is the lightest superparticle. This scenario is general enough so that our main conclusions can hold in other supersymmetric scenarios.
A detailed description of this study can be found in \cite{Arbey:2017eos}.

\section{Method}
For this study, 20 million pMSSM points have been generated in a random flat scan with SOFTSUSY \cite{softsusy}, taking $1<\tan \beta <60$, sfermion masses up to 3 TeV, gaugino mass parameters between -3 and 3 TeV, $50<M_{A}<2000$ GeV, and trilinear couplings between $-10$ and 10 TeV. SuperIso Relic \cite{superisorelic} and MicrOMEGAs \cite{micromegas} are used to compute dark matter observables. As the event generations and detector simulations needed for the LHC analysis is computationally very time consuming, we excluded from the sample of points, those in disagreement with the light Higgs mass constraint, flavour physics, and LEP and Tevatron limits, as well as the upper bound of the relic density \cite{Planck}. The remaining sample of points is then composed of 63\% of wino-like neutralinos, 32\% of higgsino-like neutralinos, 3\% of bino-like neutralinos and 2\% of mixed-state neutralinos (a neutralino is said to be of a certain type if the corresponding element in the neutralino mixing matrix is larger than 90\%). The small number of bino-like neutralinos is due to the upper bound of the relic density, as binos have generally small annihilation cross-sections and hence large relic densities.

\begin{figure}[t!]
\begin{center}
\begin{tabular}{cc}
\underline{Direct detection}&\underline{Indirect detection}\\
\includegraphics[width=0.4\columnwidth]{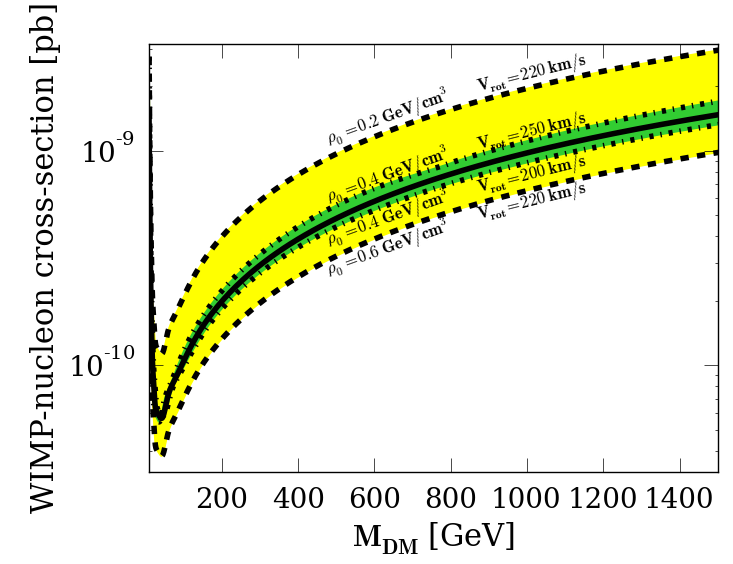}&
\includegraphics[width=0.45\textwidth]{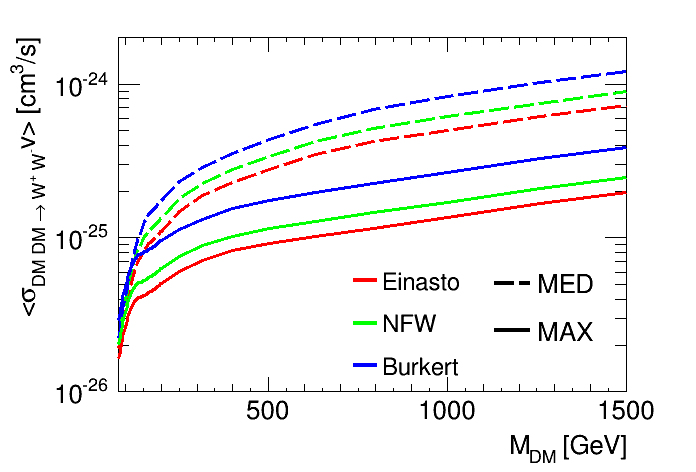}\\
\end{tabular}
\caption{\textbf{Left}: XENON1T 90\% C.L. spin-independent \textsc{WIMP}-nucleon cross section upper limit for $\rho_{0}=0.4$ GeV/cm$^{3}$ and $v_{rot}=220$ km/s (black plain line). Uncertainties on these values are shown by varying independently the DM local density (yellow band) and the disc rotation velocity (green band). \textbf{Right}: 95\% C.L. upper limit of the neutralino annihilation cross section into $W^{+}W^{-}$ derived from the AMS-02 antiproton data. \vspace*{-0.3cm}}
\label{fig:pbar_limits}
\end{center}
\end{figure}

We explored the impact of astrophysical uncertainties on the exclusions of our model points using direct and indirect detection experimental limits. For the direct detection, we used the XENON1T upper limit on the spin-independent DM-nucleon scattering cross-section~\cite{Aprile:2017iyp}. This limit strongly depends on the value of the local DM density and on the DM velocity profile in the Earth rest frame. Assuming a Standard Halo Model, the galactic disc rotation velocity is the main parameter of the velocity profile. In this work, we used three typical values $v_{rot}=$ 200, 220 and 250 km/s. As for the local DM density, we considered that it lies between 0.2 and 0.6 GeV/cm$^3$ and chose three benchmark values $\rho_{0}=$ 0.2, 0.4 and 0.6 GeV/cm$^3 $. To assess the impact of these parameters, we have rescaled the XENON1T limit according to the different parameter values. The results are shown in figure \ref{fig:pbar_limits}. It appears clearly that the local DM density value is the dominant source of uncertainty in the direct detection limit as it can shift the limit by a factor 2.
Concerning indirect detection, we mainly focused on the constraints coming from AMS-02 antiproton data \cite{AMS02}. Following the procedure described in \cite{atlast}, we derived limits on the neutralino annihilation cross section times velocity, using three different DM halo profiles: Einasto, NFW and Burkert \cite{einasto, NFW, burkert}, and the benchmark models \MED and \MAX describing antiproton propagation through the galactic medium. The \MED ({\sc Max}) model provides a conservative (stringent) bound on the dark matter antiproton signal. 
The result is shown in figure~\ref{fig:pbar_limits} for the neutralino annihilating into $W^{+}W^{-}$, but similar limits were also calculated for the other channels. It can be seen that the antiproton propagation is the main source of uncertainty, changing the limits by a factor 4. The most conservative limit corresponds to the \MED model combined with the Burkert profile, while the most stringent case corresponds to the \MAX parameters with the Einasto profile. For comparison, we also performed a combined analysis of the 19 confirmed dwarf galaxies observed by Fermi-LAT \cite{Fermi-LAT:2016uux}. Based on this analysis of the astrophysical uncertainties, we defined three different scenarios for the astrophysical parameters: a \emph{conservative} case, with \MED propagation model, Burkert profile and $\rho_{0}=0.2$ GeV/cm$^3$, a \emph{standard} case using Fermi-LAT constraints and $\rho_{0}=0.4$ GeV/cm$^3$, and a \emph{stringent} case with Einasto \MAX and $\rho_{0}=0.6$ GeV/cm$^3$.

In addition, we considered limits from collider searches taking into account the LHC 13 TeV results. In particular, we considered monojet and SUSY direct searches, light Higgs signal strength constraints, as well as heavy Higgs decay constraints $H/A \to \tau \tau$ (we refer the reader to \cite{Arbey:2017eos} for more details).

\section{Constraints in the pMSSM}

\begin{figure}[t!]
\begin{tabular}{cc}
\includegraphics[width=0.45\textwidth]{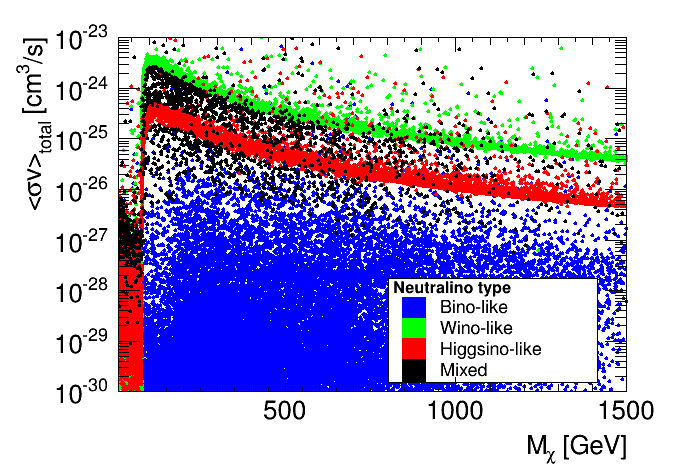} 

&
\includegraphics[width=0.45\textwidth]{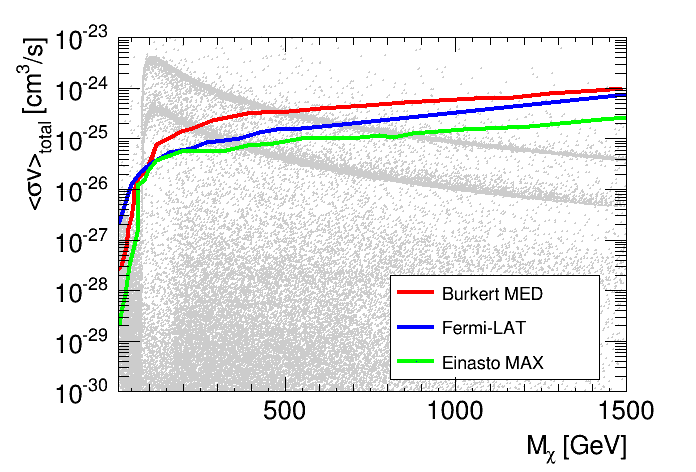} 

\end{tabular}
\caption{\textbf{Left}: Total annihilation cross section as a function of the neutralino 1 mass for the different neutralino types. \textbf{Right}: Points excluded by Fermi-LAT gamma ray and AMS-02 antiproton data in the total annihilation cross section vs. neutralino 1 mass parameter plane. The points above the red line are excluded by AMS-02 data in the conservative case with Burkert profile and \MED propagation model, above the blue line by the Fermi-LAT data, and above the green line by AMS-02 data in the stringent case with Einasto profile and \MAX propagation model.}
\label{indirect}
\end{figure}

\begin{figure}[t!]
\begin{center}
\begin{tabular}{ccc}
  \includegraphics[width=.33\linewidth]{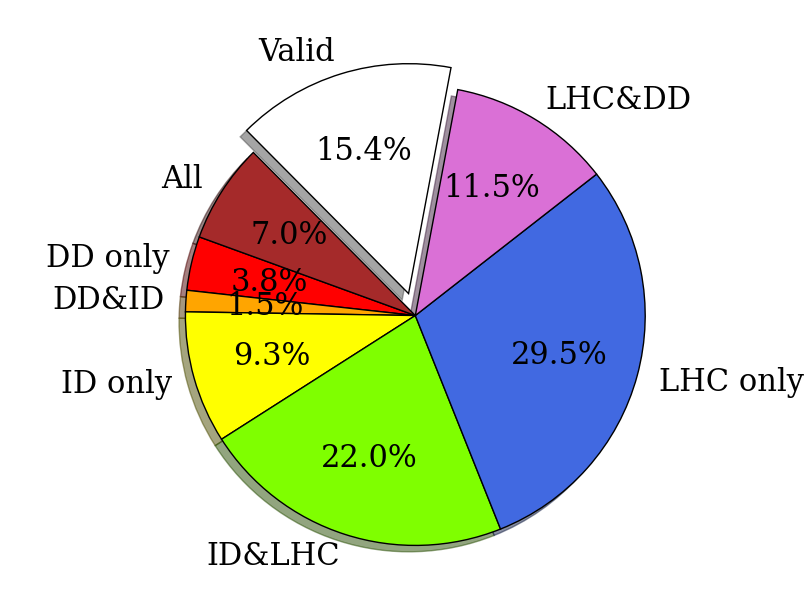}&
  \includegraphics[width=.33\linewidth]{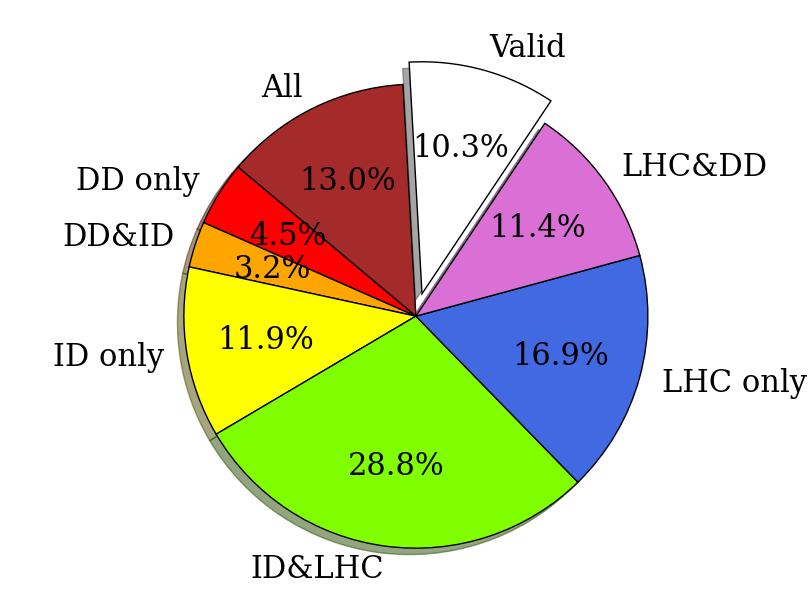}&
  \includegraphics[width=.33\linewidth]{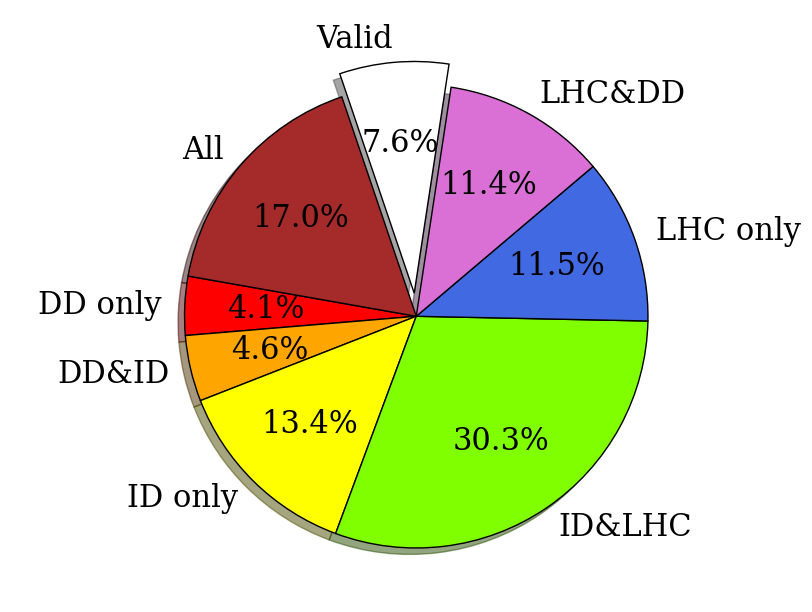}\\
   CONSERVATIVE&STANDARD&STRINGENT\\
   \end{tabular}
 \caption{Fraction of the pMSSM points satisfying the light Higgs mass, relic density, LEP and flavour constraints excluded by direct and indirect detections and LHC constraints.\label{pie:allDMLHC}}
 \end{center}
\end{figure}

The LHC search results exclude a substantial fraction of our sample of points ($\approx 70 \%$), but direct and indirect searches provide complementary constraints. Almost all the mixed-state neutralinos are excluded by direct detection constraints, independently of the astrophysical uncertainties. Next, higgsino-like neutralinos are the most excluded type of neutralinos. All types considered together, between 5 and 9\% of our points are excluded by direct detection, depending on the chosen local DM density value, while not being probed by the LHC (see figure~\ref{pie:allDMLHC}). Indirect detection constraints present also interesting features. As can be seen in figure \ref{indirect}, winos and higgsinos form two separate strips with large annihilation cross sections for masses above 80 GeV in the ($M_{\chi}, \left< \sigma v \right>$) parameter plane. Winos annihilate mostly into $W^{+}W^{-}$, and higgsinos into $W^{+}W^{-}$ and $ZZ$. Imposing the constraints from antiproton data in the conservative case Burkert {\sc Med}, we excluded points from the wino strip up to 500 GeV, while in the stringent case Einasto {\sc Max}, we excluded winos up to $\approx 1$ TeV. Fermi-LAT upper limit lies between those two extreme cases, excluding winos between 80 and 700 GeV and also a significant fraction of higgsinos (see figure \ref{indirect}). In addition to the LHC constraints, indirect detection allowed us to exclude between 11 and 18\% of the points, depending on the astrophysical parameter values. The indirect detection contribution can then be substantial, yet it still suffers from large astrophysical uncertainties. Indirect detection is particularly important since it can exclude some compressed scenarios, even in the most conservative case, that are not probed by the LHC.

\section{Conclusions}

We studied the implications of dark matter searches, together with collider constraints, on the
pMSSM with neutralino dark matter focusing on the consequences of the related uncertainties. The latest LHC constraints exclude 70\% of our sample of points. We showed that direct detection constraints from XENON1T exclude a robust fraction of our points, essentially with a higgsino-like neutralino 1. In addition to the LHC, between 5 and 9\% of the points are excluded depending on the value of the local DM density. Indirect detection constraints on the other hand are more sensitive to the astrophysical uncertainties, and depending on the astrophysical scenario can exclude an additional fraction of 11 to 18\% of the points. Indirect detection can then have a significant role, especially when considering the prospects of future experiments such as CTA \cite{CTA}, but before improvements in our knowledge of the galactic center, it will not lead to solid conclusions in terms of MSSM constraints.

In conclusion, when applying constraints from direct detection experiments on the MSSM, an intrinsic uncertainty of a factor 3 in scattering cross-sections needs to be considered to take into account the uncertainties on the local dark matter density and velocity. Similarly, when considering indirect detection experiment limits in the MSSM, for the gamma-ray spectra a variation of about 30\% in the annihilation cross-sections is possible due to the uncertainties from dark matter profiles in dwarf spheroidals, and for the antiproton limits one order of magnitude mainly due to the propagation models.


\begin{thebibliography}{99}

\bibitem{Arbey:2017eos}
  A.~Arbey, M.~Boudaud, F.~Mahmoudi and G.~Robbins,
  arXiv:1707.00426 [hep-ph].
  
\bibitem{softsusy} B. C. Allanach, Comput. Phys. Commun. \textbf{143} (2002) 305 [hep-ph/0104145].
  
\bibitem{superisorelic}  A.  Arbey  and  F.  Mahmoudi,  Comput.  Phys.  Commun.
\textbf{181} (2010) 1277 [arXiv:0906.0369].

\bibitem{micromegas} G.~Belanger, F.~Boudjema, A.~Pukhov and A.~Semenov, Comput. Phys. Commun. {\bf 185} (2014) 960 [arXiv:1305.0237].

\bibitem{Planck} P.~A.~R.~Ade {\it et al.} [Planck Collaboration],
  Astron.\ Astrophys.\  {\bf 594} (2016) A13
  [arXiv:1502.01589].

\bibitem{Aprile:2017iyp}
  E.~Aprile {\it et al.} [XENON Collaboration],
  arXiv:1705.06655 [astro-ph.CO].

\bibitem{AMS02} M.~Aguilar {\it et al.}, Phys. Rev. Lett.\ {\bf 117} (2016) 9. 

\bibitem{atlast}
  G.~Giesen, M.~Boudaud, Y.~G\'enolini, V.~Poulin, M.~Cirelli, P.~Salati and P.~D.~Serpico,
  JCAP {\bf 1509} (2015) 023
  [arXiv:1504.04276].
  
\bibitem{einasto}
  R.~Catena and P.~Ullio,
  JCAP {\bf 1008} (2010) 004
  [arXiv:0907.0018].

\bibitem{NFW} 
P.~J.~McMillan,
  MNRAS {\bf 465} (2017) no.76, 94 [arXiv:1608.00971].


\bibitem{burkert}
  F.~Nesti and P.~Salucci,
  JCAP {\bf 1307} (2013) 016
  [arXiv:1304.5127].
  

\bibitem{Fermi-LAT:2016uux}
  A.~Albert {\it et al.} [Fermi-LAT and DES Collaborations],
  Astrophys.\ J.\  {\bf 834} (2017) no.2,  110
  [arXiv:1611.03184].

  
\bibitem{CTA}
  J.~Carr {\it et al.} [CTA Collaboration],
  PoS ICRC {\bf 2015} (2016) 1203
  [arXiv:1508.06128].

\end{thebibliography}
\end{document}